\documentclass[aps,prx,twocolumn,showpacs,superscriptaddress,floatfix]{revtex4-1}

\usepackage{amsmath,amssymb}
\usepackage{graphicx}
\usepackage{color}
\usepackage{rotating}
\usepackage{bm}
\usepackage{setspace}
\usepackage{hyperref}
\usepackage{multirow}

\hypersetup{
colorlinks=true,
urlcolor= blue,
citecolor=blue,
linkcolor= blue,
bookmarks=true,
bookmarksopen=false,
}

\DeclareUnicodeCharacter{2212}{-}

\renewcommand{\vec}[1]{\bm{#1}}

\begin{document}

\title{Planar Hall effect with sixfold oscillations in a Dirac antiperovskite}

\author{D. Huang}
\email{d.huang@fkf.mpg.de}
\affiliation{Max Planck Institute for Solid State Research, 70569 Stuttgart, Germany}
\author{H. Nakamura}
\email{hnakamur@uark.edu}
\affiliation{Department of Physics, University of Arkansas, Fayetteville, Arkansas 72701, USA}
\author{H. Takagi}
\email{h.takagi@fkf.mpg.de}
\affiliation{Max Planck Institute for Solid State Research, 70569 Stuttgart, Germany}
\affiliation{Institute for Functional Matter and Quantum Technologies, University of Stuttgart, 70569 Stuttgart, Germany}
\affiliation{Department of Physics, University of Tokyo, 113-0033 Tokyo, Japan}

\date{\today}

\begin{abstract}
The planar Hall effect (PHE), wherein a rotating magnetic field in the plane of a sample induces oscillating transverse voltage, has recently garnered attention in a wide range of topological metals and insulators. The observed twofold oscillations in $\rho_{yx}$ as the magnetic field completes one rotation are the result of chiral, orbital and/or spin effects. The antiperovskites $A_3B$O ($A$ = Ca, Sr, Ba; $B$ = Sn, Pb) are topological crystalline insulators whose low-energy excitations are described by a generalized Dirac equation for fermions with total angular momentum $J = 3/2$. We report unusual sixfold oscillations in the PHE of Sr$_3$SnO, which persisted nearly up to room temperature. Multiple harmonics (twofold, fourfold and sixfold), which exhibited distinct field and temperature dependencies, were detected in $\rho_{xx}$ and $\rho_{yx}$. These observations are more diverse than those in other Dirac and Weyl semimetals and point to a richer interplay of microscopic processes underlying the PHE in the antiperovskites.
\end{abstract}


\maketitle


\section{Introduction}

The planar Hall effect (PHE), in contrast to the ordinary Hall effect (OHE), is defined as the transverse voltage that arises when an \textit{in-plane} magnetic field is applied. First reported in ferromagnetic metals and semiconductors~\cite{Koch_ZN_1955, Goldberg_PR_1954}, the PHE has recently gained traction in a growing number of topological metals and insulators as a tool to probe their relativistic quasiparticles~\cite{Burkov_PRB_2017, Nandy_PRL_2017, Li_PRB_2018_Wang, Liang_PRX_2018, Kumar_PRB_2018, Chen_PRB_2018, Singha_PRB_2018, Li_PRB_2018, Wu_PRB_2018, Yang_PRM_2019, Liu_PRB_2019, Liang_AIPAdv_2019, Meng_JPCM_2019, Li_PRB_2019, Li_JAP_2020, Yang_PRR_2020, Taskin_NatComm_2017, Petrushevsky_PRB_2017, Zhou_PRB_2019}. The interplay of chiral, orbital and spin degrees of freedom that governs the PHE in these materials is rich and remains subject to active investigation.

\begin{figure}
\includegraphics[scale=1]{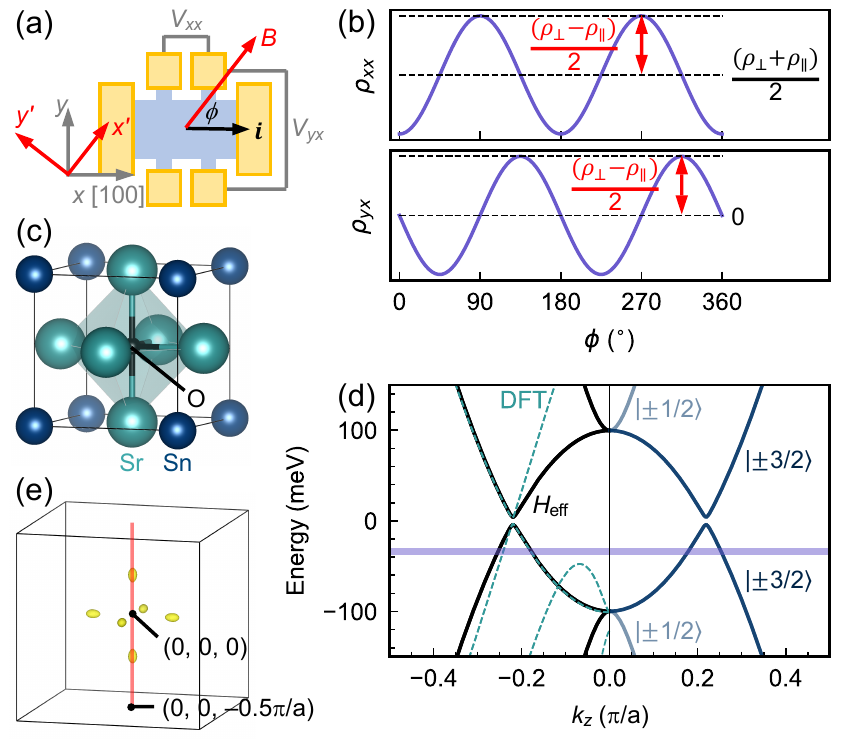}
\caption{(a) Device and measurement geometry. (b) Noncrystalline AMR ($\rho_{xx}$) and PHE ($\rho_{yx}$) as predicted by Eqs.~(\ref{Eqrhoxx}) and~(\ref{Eqrhoyx}), assuming that $\rho_{\perp}$ $>$ $\rho_{\parallel}$. (c) Crystal structure of Sr$_3$SnO. (d) Approximate band structure as described by a $\vec{k} \cdot \vec{p}$ Hamiltonian ($H_{\textrm{eff}}$). The bands along the $k_z$ axis are Kramers doublets and $j_z$ eigenstates $|\pm\tfrac{1}{2}\rangle$ and $|\pm\tfrac{3}{2}\rangle$ (right half of the plot). Bands calculated from density functional theory (DFT) are overlaid (left half of the plot). The purple line denotes the doping level corresponding to the carrier density of the measured films. (e) Fermi surface derived from $H_{\textrm{eff}}$. Only a portion of the BZ, where pockets exist, is shown. The red line denotes the cut along $k_z$ plotted in (d).}
\label{Fig1}
\end{figure}

At a phenomenological level, the PHE is driven by a magnetic-field-induced rotation of the principal axes of the resistivity tensor~\cite{Taskin_NatComm_2017}. Suppose that an in-plane magnetic field $\vec{B}$ generates a resistive anisotropy whose principal axis $x'$ is locked along the direction of $\vec{B}$ [Fig.~\ref{Fig1}(a)]. Then
\begin{equation}
\begin{pmatrix} E_{x'} \\ E_{y'} \end{pmatrix} = \begin{pmatrix} \rho_{\parallel} & 0 \\ 0 & \rho_{\perp} \end{pmatrix} \begin{pmatrix} i_{x'} \\ i_{y'} \end{pmatrix}, 
\label{Eqtrans}
\end{equation}
where $\vec{E}$ is the electric field, $\vec{i}$ is the current density and $\rho_{\parallel}$ ($\rho_{\perp}$) is the magnetoresistivity when the current runs parallel (perpendicular) to $\vec{B}$. For an arbitrary in-plane angle $\phi$ between $\vec{B}$ and $\vec{i}$, the resistivity tensor in the sample frame ($x$ along $\vec{i}$) can be obtained from a coordinate transformation:
\begin{multline}
\begin{pmatrix} E_{x} \\ E_{y} \end{pmatrix} = \begin{pmatrix} \cos\phi & -\sin\phi  \\ \sin\phi & \cos\phi \end{pmatrix} \begin{pmatrix} \rho_{\parallel} & 0 \\ 0 & \rho_{\perp} \end{pmatrix} \times \\ 
\begin{pmatrix} \cos\phi & \sin\phi  \\ -\sin\phi & \cos\phi \end{pmatrix} \begin{pmatrix} i_x \\ i_y \end{pmatrix}.
\label{Eqtrans2}
\end{multline}
By definition, $i_y$ = 0, and we extract $\rho_{xx}$ and $\rho_{yx}$:
\begin{equation}
\rho_{xx} = \frac{(\rho_{\perp}+\rho_{\parallel})}{2} - \frac{(\rho_{\perp} - \rho_{\parallel})}{2}\cos 2\phi, 
\label{Eqrhoxx}
\end{equation}
\begin{equation}
\rho_{yx} = - \frac{(\rho_{\perp} - \rho_{\parallel})}{2}\sin2\phi.
\label{Eqrhoyx}
\end{equation}
Here, $\rho_{xx}$ and $\rho_{yx}$ exhibit oscillations with two periods in a full rotation, 45$^{\circ}$ relative phase shift and equal amplitudes proportional to the difference between $\rho_{\perp}$ and $\rho_{\parallel}$ [Fig.~\ref{Fig1}(b)]. By convention, Eq.~(\ref{Eqrhoxx}) is called anisotropic magnetoresistance (AMR) and Eq.~(\ref{Eqrhoyx}) is called the PHE. From the derivation, we observe that $\rho_{yx}$ arises as an off-diagonal element in the sample frame after the coordinate transformation from the magnetic-field frame. Unlike the case of the OHE for out-of-plane fields, the resistivity tensor is symmetric and the resulting PHE is even in field.

The situation is significantly enriched when we consider the additional crystalline anisotropy introduced by the underlying lattice. Equations~(\ref{Eqrhoxx}) and~(\ref{Eqrhoyx}) are no longer exact, and the lattice symmetry dictates which higher harmonics in $\rho_{xx}$ and $\rho_{yx}$ beyond the twofold oscillations are permitted. For example, in a crystal with an in-plane square lattice, up to a sixth-order expansion in $\phi$, twofold, fourfold and sixfold oscillations are allowed in $\rho_{xx}$, whereas only twofold and sixfold oscillations are allowed in $\rho_{yx}$~\cite{Rout_PRB_2017}. Fourfold and sixfold oscillations in $\rho_{xx}$ have been observed in a number of two-dimensional (2D) oxide interfaces with square, orthorhombic or hexagonal lattices~\cite{Ngai_PRB_2011, Annadi_PRB_2013, Miao_APL_2016, Ma_PRB_2017, Rout_PRB_2017, Wadehra_NatComm_2020}. However, absent of an obvious reason, the corresponding higher harmonics in $\rho_{yx}$ were not observed, including the sixfold oscillations purported to be allowed in a 2D square lattice.

The antiperovskites $A_3B$O ($A$ = Ca, Sr, Ba; $B$ = Sn, Pb) are a unique family of compounds with unusual chemistry and electronic properties. Figure~\ref{Fig1}(c) depicts the crystal structure of Sr$_3$SnO, wherein the typical positions of the cations and anions in a perovskite are reversed. The low-energy spectrum can be described by a Dirac equation generalized to fermions with total angular momentum $J$ = 3/2~\cite{Hsieh_PRB_2014}:
\begin{equation}
H_{\textrm{eff}}(\vec{k}) = m\tau_z + v_1\tau_x\vec{k}\cdot\vec{J} + v_2\tau_x\vec{k}\cdot\vec{\tilde{J}}.
\label{EqDirac}
\end{equation}
Here, $\vec{\tau}$ is the set of Pauli matrices with $\tau_z = \pm 1$ labeling the orbitals of the valence and conduction bands, $\vec{J}$ is the set of spin-3/2 matrices and $\vec{\tilde{J}}$ is a linear combination of $\vec{J}$ and $\vec{J}^3$ that shares the same transformation properties as $\vec{k}$ under the cubic point group. Equation~(\ref{EqDirac}) describes the inversion of two $J$ = 3/2 quartets near the Brillouin zone (BZ) center, resulting in three-dimensional (3D), slightly gapped Dirac nodes along six equivalent $\Gamma$-$X$ cuts~\cite{Kariyado_JPSJ_2011, Kariyado_JPSJ_2012, Kariyado_PRM_2017} [Figs.~\ref{Fig1}(d) and~\ref{Fig1}(e)]. Microscopically, these states are composed of Sr 4$d$ and Sn 5$p$ orbitals. The latter orbitals experience strong spin-orbit coupling and come in the linear combinations $-(|p_x \uparrow\rangle + i|p_y \uparrow)/\sqrt{2}$ and $(|p_x \downarrow\rangle - i|p_y \downarrow)/\sqrt{2}$, where $\uparrow$ and $\downarrow$ denote the spin, corresponding to the eigenstates $|j_z = +\tfrac{3}{2}\rangle$ and $|j_z = -\tfrac{3}{2}\rangle$~\cite{Kariyado_JPSJ_2011}. Sr$_3$SnO is also a candidate topological crystalline insulator hosting type-I and -II Dirac surface states~\cite{Hsieh_PRB_2014, Chiu_PRB_2017}, as well as a higher-order topological insulator hosting hinge states under broken symmetry~\cite{Fang_PRB_2020}. 

Previous experiments of Sr$_3$SnO have probed its 3D Dirac carriers~\cite{Kitagawa_PRB_2018,  Rost_APLM_2019, Ma_AM_2020}, observed superconductivity upon hole doping~\cite{Oudah_NatComm_2016} (with possible implications for higher spin Cooper pairing~\cite{Kawakami_PRX_2018}) and detected weak antilocalization (WAL) due to spin-orbital entanglement~\cite{Nakamura_NatComm_2020}. Here, we measured the in-plane-field magnetoresistance of thin films of Sr$_3$SnO(001) grown by molecular beam epitaxy (MBE) and found them to be a candidate system with an in-plane square lattice exhibiting sixfold oscillations in the PHE. More specifically, the AMR showed twofold and fourfold oscillations, while the PHE showed twofold and sixfold oscillations. Remarkably, the sixfold oscillations persisted up to the highest temperature measured ($T$ = 250 K). The different harmonics in the AMR and PHE also behaved distinctly with respect to magnetic field and temperature, pointing to a complex interplay of underlying microscopic mechanisms. We speculate on the possible role of the spin-orbital-entangled $J$ = 3/2 fermions in some of these observations.

\section{Methods}

Thin films of Sr$_3$SnO were deposited on yttria-stabilized ZrO$_2$ in an Eiko MBE chamber with base pressure in the range of mid-10$^{-10}$ Torr. Detailed growth procedures, as well as film characterization via x-ray diffraction (XRD), low-energy electron diffraction and x-ray photoelectron spectroscopy, can be found in Refs.~\cite{Huang_PRM_2019, Nakamura_NatComm_2020}.

Due to the extreme chemical reactivity of the antiperovskites, post-deposition lithography of the films is challenging. Instead, during deposition, we used sapphire masks with millimeter-sized cutouts shaped as Hall bars to obtain films with well-defined geometries (width $W$ = 2 mm, aspect ratio $W/L$ = 2). For each deposition, four substrates were used, since each sample could only be used for one measurement. After the MBE growth, the films were transported via a vacuum suitcase to an Ar glove box furnished with a metal evaporator. The films dedicated to thickness calibration (Dektak stylus profilometer) or XRD were covered with Au to prevent degradation in ambient environment. For the films dedicated to transport, we evaporated Au contact pads, covered the films with grease (Apiezon N) and attached them to rotation sample pucks using Au wires and Ag epoxy (EPO-TEK E4110, low-temperature cure), all within the glove box. Two distinct pucks were used to allow in-plane and out-of-plane rotation in a magnetic field. The films were then rapidly transported to a physical property measurement system with a 14 T magnet (Quantum Design). The base temperature for the rotation insert was 1.8 K, and the highest temperature that could be measured before the antiperovskite films showed signs of degradation was 250 K. 

Our Eiko MBE chamber is attached to a customized Unisoku scanning tunneling microscope (STM) that operates at room temperature, with base pressure in the range of mid-10$^{-10}$ Torr. Films for imaging were transferred \textit{in situ} from the MBE to the STM. W tips were sharpened via electron-beam bombardment. Atomic-resolution imaging was confirmed on highly oriented pyrolitic graphite.   

The DFT bands shown in Fig.~\ref{Fig1}(d) were reproduced from Ref.~\cite{Huang_PRM_2019}. Detailed parameters can be found there. The doping level corresponding to the carrier density of the films was determined by integrating the density of states. Measured concentrations of 1.9--2.6 $\times$ 10$^{19}$ holes/cm$^{3}$ correspond to a doping level of 30 meV below the valence band maximum.

Following Ref.~\cite{Hsieh_PRB_2014}, we made a substitution to the mass term in $H_{\textrm{eff}}$ [Eq.~(\ref{EqDirac})]: $m \rightarrow m + \alpha k^2$. This was deemed necessary to properly classify the topology of the system. Parameters used to reproduce the DFT band structure are $m$ = $-$0.09975 eV, $v_1$ = 0.134 eV, $v_2$ = 0.39 eV and $\alpha$ = 0.21 eV. These values result in a mirror Chern number of $n_M$ = $-2$, although identical band dispersions with $n_M$ = $+2$ are also possible with different choices of $v_1$ and $v_2$.

The crystal structure in Fig.~\ref{Fig1}(c) and the isosurface in Fig.~\ref{Fig1}(e) were visualized using VESTA~\cite{Momma_JAC_2011}.

\section{Results}

\subsection{Surface characterization}

\begin{figure}
\includegraphics[scale=1]{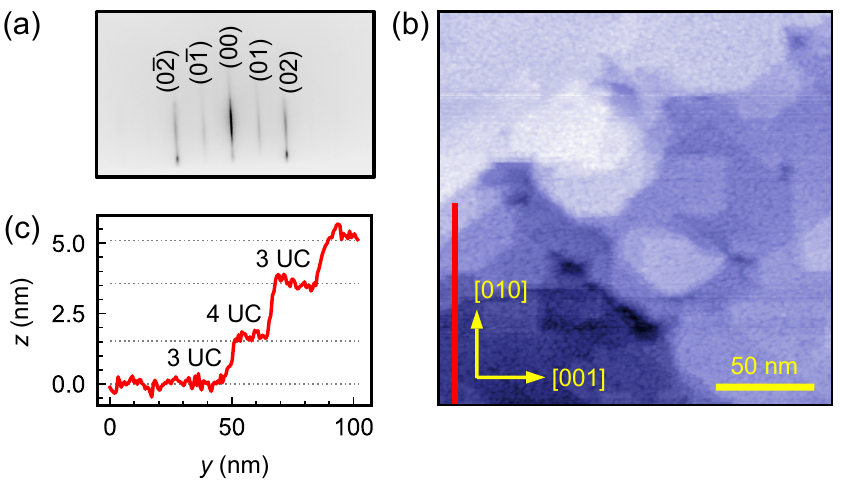}
\caption{(a) RHEED image of Sr$_3$SnO(001) taken along the [100] direction (sample SS121). Electron energy: 15 keV. (b) STM topography acquired at room temperature. Set point: $-1.75$ V, 10 pA. (c) Averaged line cut along the vertical red line in (b), which has a width of 5 pixels. The steps correspond to integer multiples of the UC (approximately 0.51 nm), as labeled in the plot.}
\label{Fig2}
\end{figure}

Figure~\ref{Fig2}(a) shows reflection high-energy electron diffraction (RHEED) patterns detected after film deposition (sample SS121), exhibiting streaks corresponding to a $1 \times 1$ lattice of Sr$_3$SnO(001). The same film was then imaged at room temperature via STM under ultra-high vacuum [Fig.~\ref{Fig2}(b)]. We observed islands with edges oriented preferably along the [110] and [$\bar{1}$10] crystal axes, and sometimes along [100] and [010], attesting to the epitaxial growth of the film. Line-cut analysis [Fig.~\ref{Fig2}(c)] reveals step edges with heights corresponding to integer multiples of the unit cell (UC), roughly 0.51 nm. We were unable to resolve an atomic lattice, likely due to complications arising from surface polarity and valence instability~\cite{Huang_PRM_2019}. 

\subsection{Transport}

\begin{figure}
\includegraphics[scale=1]{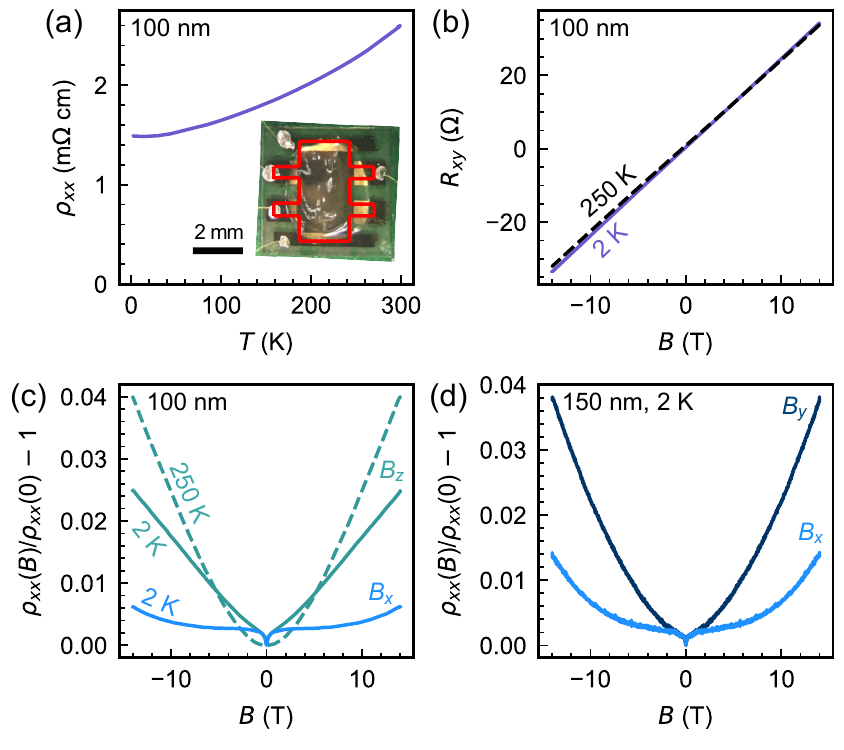}
\caption{(a) Longitudinal resistivity measured for a 100-nm-thick film (SS124) from room temperature down to 2 K. Inset: optical micrograph of the Hall-bar device. (b) OHE due to an out-of-plane magnetic field, measured at 2 K and 250 K. (c) Magnetoresistivity at 2 K and 250 K for an out-of-plane field ($B_z$) and at 2 K for an in-plane field parallel to the current ($B_x$). (d) In-plane-field magnetoresistivity at 2 K for a 150-nm-thick film (SS123). The magnetoresistivities measured under $B_x$ and $B_y$ correspond to $\rho_{\parallel}$ and $\rho_{\perp}$, respectively. The traces in (c, d) are symmetrized with respect to field.}
\label{Fig3}
\end{figure}

Figure~\ref{Fig3}(a) presents the temperature-dependent resistivity $\rho_{xx}$ of a 100-nm-thick film (SS124), which manifests semimetallic behavior. The residual resistivity ratio (RRR), defined using data points at $T$ = 297 K and 2 K, is roughly 1.7. We note that the temperature dependence and RRR are very similar to Sr$_3$SnO films measured using \textit{in-situ} four-probe transport~\cite{Ma_AM_2020}; therefore, we conclude that no significant degradation was induced by capping our films with grease. 

\begin{table*}
\setlength{\tabcolsep}{7pt}
\caption{The carrier densities were extracted from the OHE at 2 K, which then yielded the corresponding mobilities.}
\begin{tabular}{cccccc}
\hline \hline
Sample & Figures & Thickness & Carrier density & Mobility & Measurement \\
& & (nm) & ($10^{19}$ holes/cm$^{3}$) & [cm$^2$/(V$\cdot$s)] & \\
\hline
SS114 & \ref{Fig5}, \ref{Fig6} & 200 & 2.4 & 343 & Transport \\
SS119 & \ref{Fig5}, \ref{Fig6} & 50 & 1.9 & 12 & Transport \\
SS121 & \ref{Fig2} & 100 & -- & -- & STM \\
SS123 & \ref{Fig3}, \ref{Fig5}, \ref{Fig6} & 150 & 1.9 & 92 & Transport \\
SS124 & \ref{Fig3}, \ref{Fig4}, \ref{Fig5}, \ref{Fig6} & 100 & 2.6 & 162 & Transport \\
\hline \hline
\end{tabular}
\label{TSamples}
\end{table*}

Figure~\ref{Fig3}(b) displays OHE measurements in an out-of-plane field at $T$ = 2 K and 250 K. $R_{xy}$ is linear and has a positive slope corresponding to hole conduction, without any hints of mixed carriers. Minimal temperature dependence is observed. The extracted hole density and mobility at 2 K are 2.6 $\times$ 10$^{19}$ cm$^{-3}$ and 162 cm$^2$/(V$\cdot$s). The corresponding Fermi surface comprises the six Dirac pockets shown in Fig.~\ref{Fig1}(e), as well as an additional set of pockets along $\Gamma$-$M$ not captured by the effective model~\cite{Nakamura_NatComm_2020}. The parameters of all devices reported in this work are summarized in Table~\ref{TSamples}.

The magnetoresistivity in an out-of-plane field is shown in Fig.~\ref{Fig3}(c). At 250 K, $\rho_{xx}$ increases by a few percent at 14 T and displays a $B^2$ dependence characteristic of an orbital effect; however, our subsequent discussion will caution against such a straightforward interpretation. At 2 K, $\rho_{xx}$ shows a dip due to WAL below fields of 1 T. At higher fields, $\rho_{xx}$ appears to scale linearly with field, but we note that we did not consistently observe this behavior across all samples. Interestingly, the percentage change in $\rho_{xx}$ at 14 T is smaller at 2 K than at 250 K. Also shown in Fig.~\ref{Fig3}(c) is the magnetoresistivity trace at 2 K for an in-plane field, aligned along the current direction. The WAL dip remains identical, but the high-field magnetoresistivity is moderately reduced. The isotropy of the WAL attests to the film being in an electronically 3D regime; further support can be found in Ref.~\cite{Nakamura_NatComm_2020}. 

Since AMR and the PHE involve rotations of the magnetic field in the plane of the sample, we show the in-plane-field magnetoresistivity of another 150-nm-thick sample (SS123) at 2 K [Fig.~\ref{Fig3}(d)]. The in-plane field is applied parallel and perpendicular to the current, corresponding to $\rho_{\parallel}$ and $\rho_{\perp}$ in Eqs.~(\ref{Eqrhoxx}) and~(\ref{Eqrhoyx}). Again, the WAL dip remains identical, but $\rho_{\perp}$ $>$ $\rho_{\parallel}$ at higher fields.

\subsection{AMR and PHE}

\begin{figure*}
\includegraphics[scale=1]{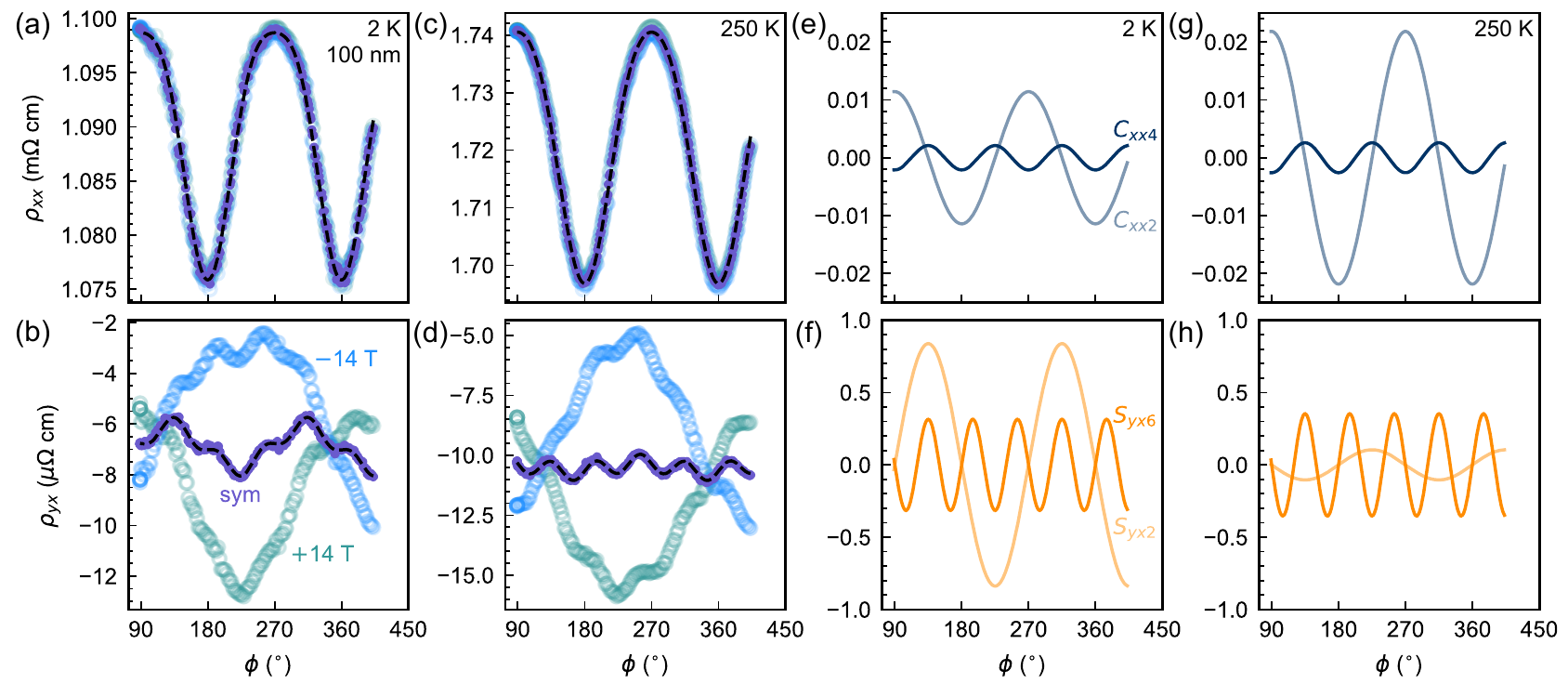}
\caption{(a, b) AMR ($\rho_{xx}$) and PHE ($\rho_{yx}$) at $T$ = 2 K. The open circles denote the raw data acquired at $B$ = $\pm$14 T, while the filled circles denote the field-symmetrized data. The overlaid dashed lines are fits to Eqs.~(\ref{Eqfit_AMR}) and~(\ref{Eqfit_PHE}) in the text. (c, d) AMR and PHE at $T$ = 250 K. (e)--(h) Decomposition of the fits in (a)--(d), respectively, to various harmonics. Film thickness is 100 nm (SS124).}
\label{Fig4}
\end{figure*}

We acquired $\rho_{xx}$ (AMR) and $\rho_{yx}$ (PHE) for the 100-nm-thick film (SS124) while applying a current along the [100] direction and rotating the device in a fixed, in-plane magnetic field of 14 T. Here, $\phi$ is defined relative to the [100] axis. Measurements at 2 K and 250 K are displayed in Figs.~\ref{Fig4}(a)--\ref{Fig4}(d). Due to wobbling in the sample puck during rotation, the device experienced a small out-of-plane component of $\vec{B}$ with changing magnitude. In the $\rho_{yx}$ channel, an out-of-plane field generates an additional contribution from the OHE, which is odd in field. This spurious contribution can be cleanly removed by symmetrizing $\rho_{yx}$ acquired at $+$14 T and $-$14 T [Figs.~\ref{Fig4}(b) and~\ref{Fig4}(d)], since the PHE itself is even in field~\cite{SM}. On the other hand, as seen in Figs.~\ref{Fig4}(a) and~\ref{Fig4}(c), $\rho_{xx}$ is nearly identical for $+$14 T and $-$14 T. 

Qualitatively, the oscillations in $\rho_{xx}$ largely obey the $\cos 2\phi$ dependence of Eq.~(\ref{Eqrhoxx}). Maxima appear at $\phi$ = $90^{\circ}$ and $270^{\circ}$ and minima appear at $\phi$ = $180^{\circ}$ and $360^{\circ}$. The negative amplitude indicates that $\rho_{\perp} > \rho_{\parallel}$. Closer inspection of Figs.~\ref{Fig4}(a) and~\ref{Fig4}(c) reveals an asymmetry between the maxima, which have a broader width, and the minima, which have a narrower width, hinting at higher harmonics. The situation is significantly more pronounced in the $\rho_{yx}$ channel. At 2 K [Fig.~\ref{Fig4}(b)], maxima appear at $\phi$ = $135^{\circ}$ and $315^{\circ}$ and minima appear at $\phi$ = $225^{\circ}$ and $405^{\circ}$, consistent with a negative $\sin 2\phi$ form. However, higher harmonics are visibly superimposed. At 250 K [Fig.~\ref{Fig4}(d)], the oscillations in $\rho_{yx}$ become predominantly sixfold, with maxima spaced 60$^{\circ}$ apart.

Motivated by the allowed harmonics for a 2D square lattice~\cite{Rout_PRB_2017}, we employed the following minimal model to fit the observed AMR and PHE:
\begin{equation}
\rho_{xx} = C_{xx0} + C_{xx2} \cos 2\phi + C_{xx4} \cos 4\phi,
\label{Eqfit_AMR}
\end{equation}
\begin{equation}
\rho_{yx} = S_{yx2} \sin 2\phi + S_{yx6} \sin 6\phi.
\label{Eqfit_PHE}
\end{equation}
Figures~\ref{Fig4}(e)--\ref{Fig4}(h) show the extracted twofold and fourfold oscillations in $\rho_{xx}$ and twofold and sixfold oscillations in $\rho_{yx}$. We observe that $C_{xx2}$ and $S_{yx2}$ have different magnitudes at 2 K and 250 K, but $C_{xx4}$ and $S_{yx6}$ are constant. Due to small random misalignments in the transverse voltage contacts, there is some admixture of $\rho_{xx}$ in $\rho_{yx}$. Hence, we supplement Eq.~(\ref{Eqfit_PHE}) with the terms $C_{yx0}$ and $C_{yx2} \cos 2\phi$ in order to fit $\rho_{yx}$, but they are not intrinsic to the PHE.

\subsection{Field, temperature and thickness dependencies}

\begin{figure*}
\includegraphics[scale=1]{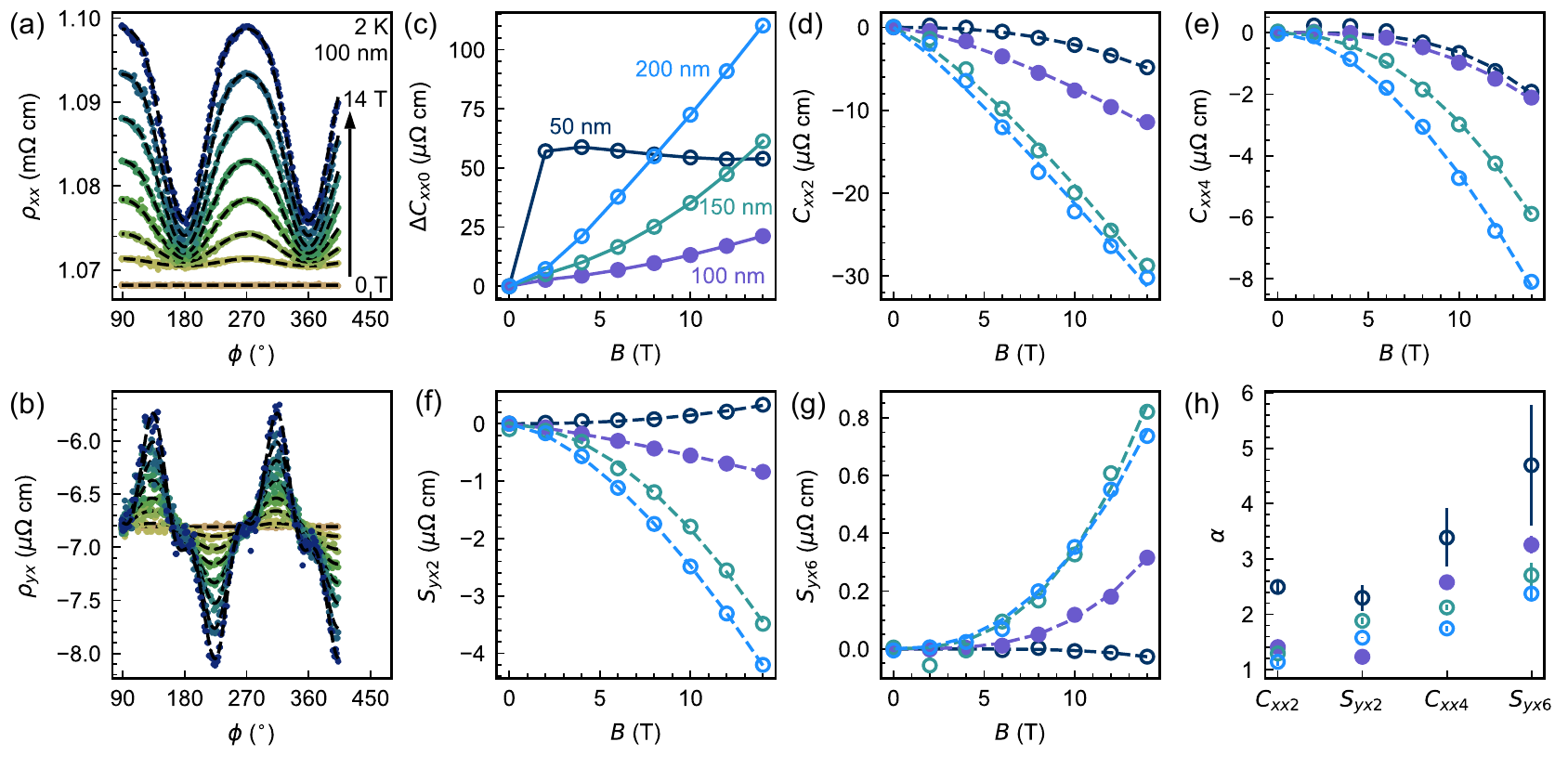}
\caption{(a, b) $\rho_{xx}$ and $\rho_{yx}$ acquired in 2 T intervals from 0 T to 14 T (direction of arrow indicates increasing field). Fits to Eqs.~(\ref{Eqfit_AMR}) and~(\ref{Eqfit_PHE}) are overlaid as dashed lines. $T$ = 2 K, film thickness is 100 nm (SS124). (c)--(g) Field dependencies of the $C_{xx0}$, $C_{xx2}$, $C_{xx4}$, $S_{yx2}$ and $S_{yx6}$ amplitudes extracted from (a, b) (filled circles). Data from three additional films with different thicknesses (SS114, SS119, SS123) are also plotted (open circles). In (c), $\Delta C_{xx0} = C_{xx0}(B) - C_{xx0}(0)$. In (d)--(g), fits to power laws of the form $B^{\alpha}$ are overlaid as dashed lines. (h) Fit values of $\alpha$ vs. harmonic for the different films.}
\label{Fig5}
\end{figure*}

We next examined the field and temperature dependencies of the AMR and PHE for a total of four films with similar hole concentrations but varying thicknesses of 50 nm (SS119), 100 nm (SS124), 150 nm (SS123) and 200 nm (SS114). Figures~\ref{Fig5}(a) and~\ref{Fig5}(b) track the evolution of the AMR and PHE at 2 K as the field is increased from 0 T to 14 T. Again, the data were symmetrized with respect to field. The corresponding evolutions of $C_{xx0}$, $C_{xx2}$, $C_{xx4}$, $S_{yx2}$ and $S_{yx6}$ are plotted in Figs.~\ref{Fig5}(c)--\ref{Fig5}(g). The $C_{xx0}$ term corresponds to the average of $\rho_{\perp}$ and $\rho_{\parallel}$. A dip at zero field due to WAL is present at 2 K, and especially pronounced in the thinnest film of 50 nm. To aid comparison across the samples, we plot the change in $C_{xx0}$, defined as $\Delta C_{xx0} = C_{xx0}(B) - C_{xx0}(0)$,  in Fig.~\ref{Fig5}(c). Since the WAL dip is identical for $\rho_{\perp}$ and $\rho_{\parallel}$ [Fig.~\ref{Fig3}(d)], it does not factor into the higher harmonics. In fact, the higher harmonics all obey some power-law dependence of the form $B^{\alpha}$. The extracted values of the exponent $\alpha$ are presented in Fig.~\ref{Fig5}(h). We note that the error bar is rather large for $\alpha(S_{yx6})$ in the 50-nm-thick film, due to the smallness of $S_{yx6}$. Despite the overall variation in $\alpha$, two trends are discernable: $\alpha$ increases as the harmonic increases and as the film thickness decreases.

\begin{figure*}
\includegraphics[scale=1]{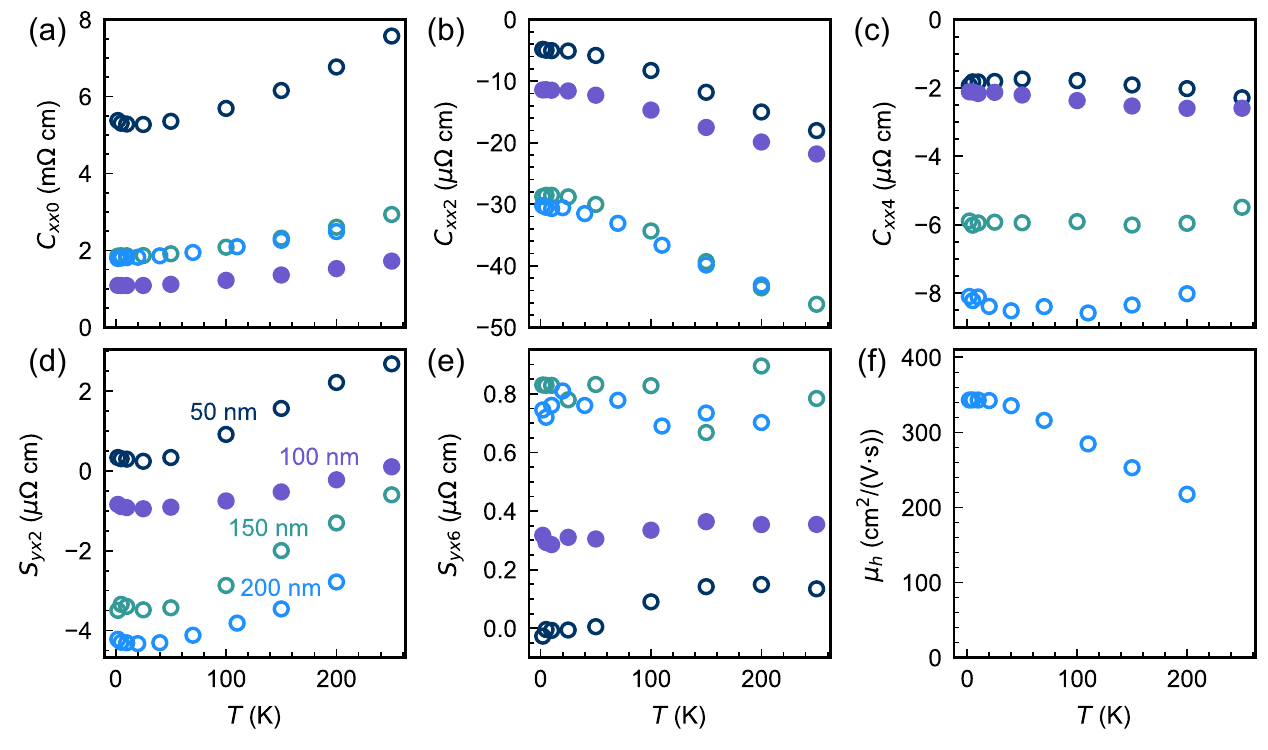}
\caption{(a)--(e) Temperature dependencies of the $C_{xx0}$, $C_{xx2}$, $C_{xx4}$, $S_{yx2}$ and $S_{yx6}$ amplitudes. Data from a total of four films with different thicknesses are shown. (f) Hole mobility corresponding to the 200-nm-thick film (SS114).}
\label{Fig6}
\end{figure*}

Figures~\ref{Fig6}(a)--\ref{Fig6}(e) track the evolutions of $C_{xx0}$, $C_{xx2}$, $C_{xx4}$, $S_{yx2}$ and $S_{yx6}$ as the temperature is raised from 2 K up to 250 K. The $C_{xx0}$ term simply mimics the temperature dependence of $\rho_{xx}$ at zero field [Fig.~\ref{Fig3}(a)]. With respect to the higher harmonics, however, we observe an unusual dichotomy between $C_{xx2}$ and $S_{yx2}$, which show clear temperature dependence, and $C_{xx4}$ and $S_{yx6}$, which are relatively independent of temperature up to 250 K. But even among the temperature-dependent harmonics, $C_{xx2}$ and $S_{yx2}$ behave dissimilarly: $C_{xx2}$ increases while $S_{yx2}$ decreases with increasing temperature. Thus, we observe harmonics that increase, decrease and remain constant with respect to temperature. This diversity is unprecedented and defies the usual expectations for semiclassical magnetoresistance, whose amplitude scales with the carrier mobility. As seen in Fig.~\ref{Fig6}(f), the hole mobility decreases with increasing temperature, a trend that is only shared by one of the harmonics, $S_{yx2}$.

From Figs.~\ref{Fig5} and~\ref{Fig6}, we observe that in general, the amplitudes of the harmonics increase as the film thickness increases. However, the 50-nm-thick sample (SS119) is anomalous in several regards. The sign of $S_{yx2}$ is flipped from negative to positive, its amplitude grows with increasing temperature, and $S_{yx6}$ is nearly suppressed below 100 K. Our previous WAL measurements established that 50-nm-thick samples lie in a quasi-2D electronic regime, based on the temperature dependence of the phase coherence length~\cite{Nakamura_NatComm_2020}. The changes in the PHE here could reflect such a 3D-2D crossover, or also the lower mobility of this thinnest sample.

\section{Discussion}

To summarize, the existence of sixfold oscillations in the PHE and the manifold behaviors of the different harmonics are distinguishing features of our experiment on Sr$_3$SnO. We emphasize that the in-plane symmetry of Sr$_3$SnO(001) is fourfold, and neither RHEED nor STM uncovered any signatures of sixfold structure. According to a symmetry analysis by Rout \textit{et al.}~\cite{Rout_PRB_2017}, the most general expansions of the AMR and PHE for a 2D square lattice are
\begin{multline}
\rho_{xx} = a_0 + a_2\cos 2\phi_B \cos 2\phi_i + b_2\sin 2\phi_B \sin 2\phi_i  \\
+ a_4\cos 4\phi_B + a_6\cos 6\phi_B \cos 2\phi_i \\
+ b_6\sin 6\phi_B \sin 2\phi_i + ...
\end{multline}
and
\begin{multline}
\rho_{yx} = -a_2\cos 2\phi_B \sin 2\phi_i + b_2\sin 2\phi_B \cos 2\phi_i  \\
- a_6\cos 6\phi_B \sin 2\phi_i + b_6\sin 6\phi_B \cos 2\phi_i + ...,
\end{multline}
where $\phi_B$ ($\phi_i$) is the angle between the magnetic field (current) and a reference crystallographic axis. In our case, the reference axis is [100], $\phi_B$ = $\phi$ and $\phi_i$ = 0, yielding 
\begin{equation}
\rho_{xx} = a_0 + a_2\cos 2\phi + a_4\cos 4\phi + a_6\cos 6\phi + ...
\label{Eqsym1}
\end{equation}
and
\begin{equation}
\rho_{yx} = b_2\sin 2\phi + b_6\sin 6\phi + ....
\label{Eqsym2}
\end{equation}
Our fitting model [Eqs.~(\ref{Eqfit_AMR}) and~(\ref{Eqfit_PHE})] thus respects the terms permitted in a 2D square lattice. From the analysis, we also conclude that the existence of $C_{xx4}$ but absence of $S_{yx4}$ is a direct consequence of the lattice symmetry. We note that Eqs.~(\ref{Eqsym1}) and~(\ref{Eqsym2}) enforce no constraints on the relative magnitudes of the harmonics. In our data, $S_{yx6}$ is clearly visible, but $C_{xx6}$ is somehow too small to extract. 

We now consider each harmonic in greater detail. $C_{xx2}$ is negative in amplitude, implying that $\rho_{\perp}$ $>$ $\rho_{\parallel}$, based on Eq.~(\ref{Eqrhoxx}). Similar $C_{xx2}$ with negative amplitude was observed in other Dirac and Weyl semimetals and ascribed to the chiral anomaly, orbital magnetoresistance, or a combination of the two~\cite{Li_PRB_2018_Wang, Liang_PRX_2018, Kumar_PRB_2018, Chen_PRB_2018, Singha_PRB_2018, Li_PRB_2018, Wu_PRB_2018, Yang_PRM_2019, Liu_PRB_2019, Liang_AIPAdv_2019, Meng_JPCM_2019, Li_PRB_2019, Li_JAP_2020}. In the chiral anomaly, the magnetoresistivity becomes negative when $\vec{B}$ and $\vec{i}$ are aligned, leading to $\rho_{\perp}$ $>$ $\rho_{\parallel}$. In orbital magnetoresistance, the Lorentz force is maximized when $\vec{B}$ and $\vec{i}$ are perpendicular, leading also to $\rho_{\perp}$ $>$ $\rho_{\parallel}$. The absence of negative longitudinal magnetoresistivity in Figs.~\ref{Fig3}(c) and~\ref{Fig3}(d) clearly favors the latter scenario. However, the temperature dependence of $C_{xx2}$ is inconsistent with a pure orbital effect. Rather than scaling with the carrier mobility and increasing at low temperatures, it decreases at low temperatures [Fig.~\ref{Fig6}(b)], similar to the magnetoresistivity in Fig.~\ref{Fig3}(c). This anomalous temperature dependence may point to a competition between orbital magnetoresistance generating negative $C_{xx2}$ and another process generating positive $C_{xx2}$ ($\rho_{\parallel}$ $>$ $\rho_{\perp}$). From Fig.~\ref{Fig6}(b), we deduce that orbital magnetoresistance dominates, resulting in an overall negative sign for $C_{xx2}$, but that the competing process with positive $C_{xx2}$ grows at low temperatures, such that the overall magnitude of $C_{xx2}$ is reduced.

A similar behavior in $C_{xx2}$ was reported in SmB$_6$, where the sign of the AMR and PHE flipped from negative to positive at low temperatures~\cite{Petrushevsky_PRB_2017, Zhou_PRB_2019}. This observation was attributed to the competition between orbital magnetoresistance in the bulk and anisotropic scattering of spin-polarized surface states. The latter produces positive $C_{xx2}$~\cite{Taskin_NatComm_2017}, but only when the sample is rotated in the plane of the magnetic field, due to some mechanisms, still under debate, involving scattering off magnetic impurities or band warping by the Zeeman effect~\cite{Zheng_PRB_2020}. To test for the possible influence of surface states, arising either from nontrivial topology or a surface reconstruction, we performed AMR measurements while rotating a sample out of the plane of the magnetic field~\cite{SM}. If surface states were responsible for the anomalous temperature dependence of $C_{xx2}$ in an in-plane field, then a different temperature dependence of $C_{xx2}$ in an out-of-plane field would be expected. The reason is that the positive $C_{xx2}$ contribution from surface states would vanish in an out-of-plane field, resulting in pure orbital magnetoresistance with negative $C_{xx2}$ whose amplitude grows at low temperatures. We found that $C_{xx2}$ behaved identically for in-plane and out-of-plane fields, with similar magnitudes and the same anomalous temperature dependence. We surmise that the competing processes must be bulk in nature, a point which we revisit.

The sizeable difference between the magnitudes of $C_{xx2}$ and $S_{yx2}$ and their contrasting temperature dependencies is another puzzle for which we presently lack a microscopic picture. On a phenomenological level, the much smaller amplitude of $S_{yx2}$ implies that $C_{xx2}$ has a sizeable crystalline contribution~\cite{Rushforth_PRL_2007}; i.e., there is a $\rho_{xx}$ contribution which modulates with the angle between $\vec{B}$ and some crystal axis, rather than the angle between $\vec{B}$ and $\vec{i}$. In this case, the principal axes of the resistivity tensor do not rotate with the magnetic field, but remain locked to the sample, such that $\rho_{yx}$, which is an off-diagonal element of the resistivity tensor in the sample frame, does not appear. The sign change in $S_{yx2}$ for the thinnest sample is another mystery. Perhaps the band structure is modified in the 2D limit, or the PHE is sensitive to the disorder potential, given that the 50-nm-thick film has the lowest mobility.

The fact that $C_{xx4}$ and $S_{yx6}$ are both relatively independent of temperature up to 250 K suggests that the two are linked in nature. The origin of their temperature independence is mysterious, but we mention in passing that phenomenologically, it is somewhat reminiscent of the linear magnetoresistance in nonstoichiometric silver chalcogenides~\cite{Xu_Nature_2007}, as well as graphene~\cite{Friedman_NL_2010} and Bi$_2$Te$_3$~\cite{Wang_PRL_2012}, which shows weak temperature dependence up to room temperature. 

Finally, it is instructive to ask what distinguishes the Dirac antiperovskite Sr$_3$SnO from other 3D Dirac and Weyl semimetals, where simpler AMR and PHE were consistently reported~\cite{Li_PRB_2018_Wang, Liang_PRX_2018, Kumar_PRB_2018, Chen_PRB_2018, Singha_PRB_2018, Li_PRB_2018, Wu_PRB_2018, Yang_PRM_2019, Liu_PRB_2019, Liang_AIPAdv_2019, Meng_JPCM_2019, Li_PRB_2019, Li_JAP_2020}. In general, the higher harmonics previously observed in $\rho_{xx}$ were associated with 2D systems~\cite{Ngai_PRB_2011, Annadi_PRB_2013, Miao_APL_2016, Ma_PRB_2017, Rout_PRB_2017, Wadehra_NatComm_2020}. (One exception is Bi(111), which showed fourfold oscillations in $\rho_{yx}$ due to its highly anisotropic pockets~\cite{Yang_PRR_2020}.) We consider the bulk spin-orbital-entangled states of Sr$_3$SnO, which at low energies may be described as $J$ = 3/2 Dirac fermions. Theoretical work by Trushin \textit{et al.}~\cite{Trushin_PRB_2009} demonstrated that AMR with positive $C_{xx2}$ is possible within a 3D, spherical Kohn-Luttinger model with $J$ = 3/2 states, due to anisotropic backscattering off spin-polarized impurities aligned along the magnetic field. While a generalization to Sr$_3$SnO with six copies of $J = 3/2$ Dirac pockets is far from straightforward, if the same mechanism could produce competing AMR with a positive $C_{xx2}$ contribution, it could explain the anomalous temperature dependence of $C_{xx2}$ in our films. Furthermore, spin-3/2 states often manifest more anisotropy than spin-1/2 states, whether in the Rashba effect, where the spin expectation value winds three times instead of once in a $2\pi$ rotation, or the Zeeman effect, where the spin splitting also modulates with the magnetic field direction~\cite{Winkler_SST_2008, Nakamura_PRL_2012}. Theoretical investigations of whether $J = 3/2$ models can produce higher harmonics in the PHE, similar in methodology to Refs.~\cite{Trushin_PRB_2009, Boudjada_PRB_2019}, would shed much light.

In conclusion, we have detected sixfold oscillations in the PHE of the Dirac antiperovskite Sr$_3$SnO. Both the AMR and PHE are unusually complex, showing multiple harmonics with distinct behaviors with respect to magnetic field, temperature and film thickness. Our work exemplifies the richness of the PHE and motivates further investigations of higher harmonics in a wide range of topological materials. Our work also exposes the need for more theoretical studies to elucidate the microscopic driving forces.

\begin{acknowledgments}

We thank Y.-h. Chan, M. Hirschmann, G. Jackeli and A. Schnyder for useful discussions and M. Dueller, C. M\"{u}hle, K. Pflaum and S. Schmid for technical support. D. H. acknowledges support from a Humboldt Research Fellowship for Postdoctoral Researchers.

\end{acknowledgments}


%

\end{document}